\documentclass[letterpaper, 10 pt, conference]{ieeeconf}  

\IEEEoverridecommandlockouts                              

\usepackage{amsmath} 

\usepackage{graphicx}

\title{\LARGE \bf
Estimation of Food Intake Quantity \\Using Inertial Signals from Smartwatches
}
\author{
    Ioannis Levi$^{1}$, Konstantinos Kyritsis$^{1,2}$, Vasileios Papapanagiotou$^{1,3}$, \\ Georgios Tsakiridis$^{1}$, and Anastasios Delopoulos$^{1, \dag}$
    \thanks{$^{1}$ Multimedia Understanding Group, Department of Electrical and Computer Engineering, Aristotle University of Thessaloniki, 54636 Greece.}
    \thanks{$^{2}$ Koios Care (BV), Filip Williotstraat 9, 2600 Antwerpen, Belgium.}
    \thanks{$^{3}$ IMPACT research group, Department of Medicine, Huddinge, Karolinska Institutet, Stockholm, Sweden.}
    \thanks{$^{\dag}$Corresponding author mail: antelopo@ece.auth.gr}
}

\usepackage{booktabs}

\begin{document}

\maketitle
\thispagestyle{empty}
\pagestyle{empty}

\begin{abstract}
Accurate monitoring of eating behavior is crucial for managing obesity and eating disorders such as bulimia nervosa. At the same time, existing methods rely on multiple and/or specialized sensors, greatly harming adherence and ultimately, the quality and continuity of data. This paper introduces a novel approach for estimating the weight of a bite, from a commercial smartwatch. Our publicly-available dataset contains smartwatch inertial data from ten participants, with manually annotated start and end times of each bite along with their corresponding weights from a smart scale, under semi-controlled conditions. The proposed method combines extracted behavioral features such as the time required to load the utensil with food, with statistical features of inertial signals, that serve as input to a Support Vector Regression model to estimate bite weights. Under a leave-one-subject-out cross-validation scheme, our approach achieves a mean absolute error (MAE) of $\mathbf{3.99}$ grams per bite. To contextualize this performance, we introduce the \textit{improvement} metric, that measures the relative MAE difference compared to a baseline model. Our method demonstrates a $\mathbf{17.41}\%$ improvement, while the adapted state-of-the-art method shows a $\mathbf{-28.89}\%$ performance against that same baseline. The results presented in this work establish the feasibility of extracting meaningful bite weight estimates from commercial smartwatch inertial sensors alone, laying the groundwork for future accessible, non-invasive dietary monitoring systems.
\end{abstract}

\section{INTRODUCTION}

Dietary monitoring is crucial for managing obesity and eating disorders such as bulimia nervosa, affecting 9\% of the global population \cite{c1}. Accurate food intake tracking benefits these conditions by improving treatment adherence and outcomes \cite{c2}, but existing methods rely on self-reporting, which can be inaccurate and burdensome, or require specialized equipment, limiting practical adoption \cite{c3}.

In addressing these monitoring challenges, various approaches have been explored in literature. Scale-based systems such as table-embedded scales \cite{c4} and smart plate systems \cite{c5}, demonstrate capability in bite detection and weight estimation through specialized hardware and machine learning techniques, but fundamentally, they require manual placement and activation. These constraints and their stationary nature render such systems impractical for free-living or on-the-go monitoring, limiting real-world adoption.

Acoustic and visual sensing approaches have demonstrated promising results in dietary monitoring. Utilizing commercial earbuds, Papapanagiotou \textit{et al.} \cite{c6} achieved notable accuracy in bite weight estimation, reporting mean absolute errors below $1$g for food-specific models and $2.1$g for general estimation across food types. In the visual domain, Akpro \textit{et al.} \cite{c7} proposed a smartphone-based system using utensils as reference points. Through image processing and metadata analysis, it estimates container dimensions and food volume, achieving $6.87\%$ relative error when combined with food type information.

Multimodal approaches have emerged as a promising direction in dietary monitoring, leveraging complementary sensing modalities to enhance estimation accuracy. In \cite{c8}, Lee Ki-Seung combined visual food recognition with Doppler signals for intake monitoring, achieving a $15.02\%$ mean absolute percentage error (MAPE). Mirtchouk \textit{et al.}, introduced a sophisticated system incorporating inertial measurement unit (IMU) data from multiple body locations, audio data from customized earbuds, and manually annotated video features \cite{c9}. Their methodology yielded a $35.4\%$ relative MAPE utilizing all data sources, demonstrating the potential benefits of sensor fusion, although their work also illustrates the inherent complexity and practical limitations of multi-device solutions in real-world applications.

Recent research focuses on solutions employing commercial devices to reduce intrusiveness and hardware complexity. Particularly, IMU sensors, commonly embedded in commercial smartwatches, have demonstrated the interesting potential in dietary monitoring applications. Kyritsis \textit{et al.} developed a deep learning framework for modeling meal micromovements using commercial smartwatch data \cite{c10}, successfully classifying fundamental eating gestures into distinct movement patterns. 
Subsequent work extended these capabilities to free-living environments, demonstrating robust bite detection, meal boundary identification, and micro-behavior analysis in both healthy individuals and those with Parkinson’s Disease \cite{c11, kyritsis2021assessment}.
While these advances establish the effectiveness of smartwatch IMU data for detecting eating episodes, their potential for quantitative measures like bite weight estimation remains unexplored.

In this work, we propose a novel method for estimating bite weight using IMU from a commercial smartwatch. For this research, a bite is defined as the event starting from the initial food gathering intent, through the final downward motion after inserting food to the mouth \cite{c10}. The proposed approach combines behavioral features extracted through signal processing of a micromovement classification model's \cite{c10} temporal predictions, with statistical characteristics of IMU, fed into an optimized Support Vector Regression (SVR) model. To evaluate our method, we collected, and made publicly-available, a dataset comprising $342$ bites across ten eating sessions, where bite events were manually annotated through video analysis and synchronized with IMU recordings and corresponding weight measurements. We assess the approach through leave-one-subject-out cross-validation (LOSO CV). To achieve this, we introduce the \textit{improvement} metric, using as a reference a statistical baseline model. Our comprehensive evaluation includes comparisons against an adapted version of literature's state-of-the-art methodology, along a deep learning fusion architecture, providing robust validation of our method's effectiveness.

\section{METHODS}

\begin{figure*}[t]
    \centering
    \includegraphics[width=0.89\textwidth]{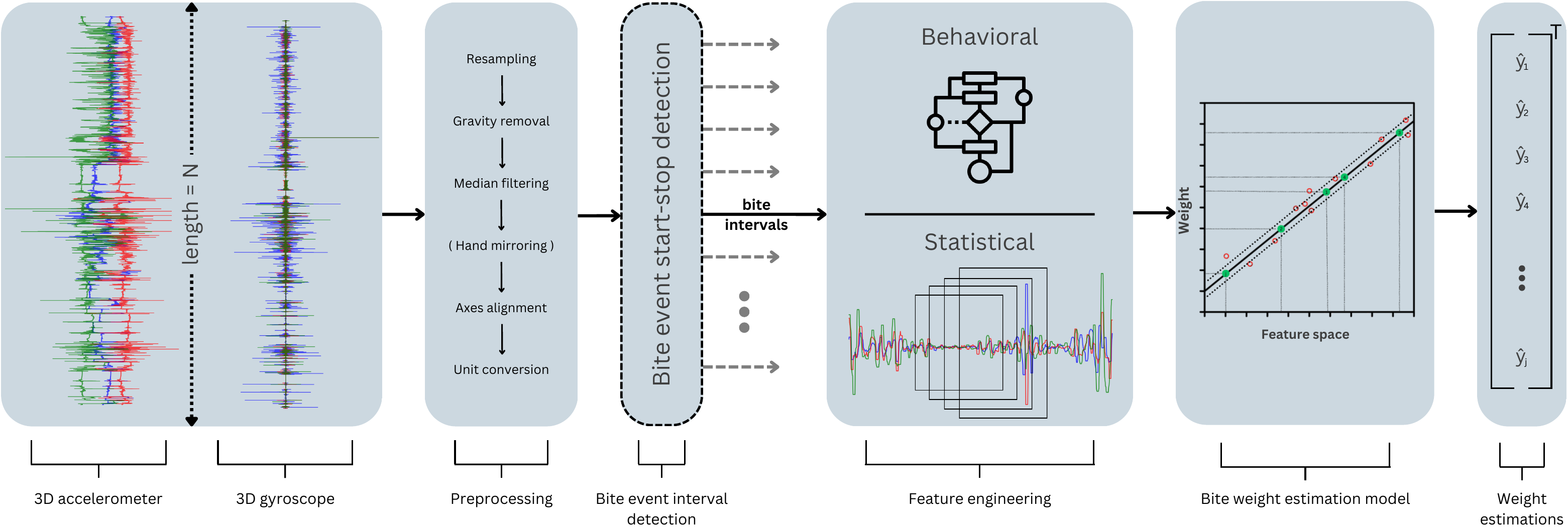}
    \caption{Overall pipeline of the proposed bite weight estimation method. From left to right: Raw IMU streams ($N = t \cdot f_s$) undergo preprocessing for signal quality optimization. Bite intervals are then identified through manual video annotation (dashed box). These intervals serve as input for automatic feature extraction, where behavioral characteristics are derived from micromovement analysis and statistical features from IMU signal processing. The resulting feature vector feeds an SVR model that generates weight estimations for each bite in the meal.}
    \label{fig:algorithm_pipeline}
\end{figure*}

\subsection{Preprocessing}

During a meal recording session, we collect synchronized 3D accelerometer and gyroscope streams, denoted as $a_x^n, a_y^n, a_z^n$ and $g_x^n, g_y^n, g_z^n$ respectively, where $n$ indexes each sample from 1 to $N$. The length $N$ of the streams is defined as $N = t \cdot f_s$, where $t$ is the meal's duration in seconds and $f_s$ the sampling rate of the IMU. These measurements are organized into a single matrix $\mathbf{M} = [\mathbf{a}_x, \mathbf{a}_y, \mathbf{a}_z, \mathbf{g}_x, \mathbf{g}_y, \mathbf{g}_z]$ of dimensions $N \times 6$.

As the first preprocessing step, we resample the inertial signals from an average of $51.84$ $(\pm 1.08)$ Hz to a constant sampling rate of $100$ Hz, through linear interpolation. Since accelerometer measurements contain both wrist movement and gravitational acceleration components, we remove the latter using a high-pass FIR filter with 501 taps and cutoff frequency of $1$ Hz, implemented with forward-backward filtering to ensure zero phase distortion. To attenuate transient signal fluctuations whilst preserving the characteristic motion patterns, a 5th-order median filter is applied, experimentally determined to provide satisfactory noise reduction. For signal consistency, all recordings collected from participants wearing the smartwatch on the left wrist are transformed to match right-wrist orientation by inverting the first, fifth and sixth channels (i.e., $a_x$, $g_y$ and $g_z$), following the hand mirroring process proposed by Kyritsis \textit{et al.} \cite{c10}.

\subsection{Feature engineering}
We propose the extraction of a total of six features ($f_{1}$ to $f_{6}$). These features are grouped into two categories; behavioral and statistical, explained in more detail below.

\subsubsection{Behavioral features}

The micromovement classification model from \cite{c10} analyzed IMU in overlapping windows of $0.2$s with $0.1$s step, producing temporal probability distributions across five distinct eating gestures: pick food (p), upward movement (u), mouth (m), downward movement (d), and no movement (n). Processing of the predictions enables the extraction of two novel behavioral features $f_{1}$ and $f_{2}$. 

The first feature, $f_{1}$, quantifies \textit{food gathering duration}. For each bite, the process initially identifies the first instance of ``m'' micromovement, denoting food insertion to mouth. Analysis of the preceding temporal sequence identifies continuous windows where the ``p'' probability exceeds $0.45$; a threshold determined through experimentation. An adaptive mechanism allows brief interruptions up to $0.2$s, integrating windows with probabilities within $[0.25, 0.45]$ when followed by strong predictions, accounting for natural variations in gathering motions.

Feature $f_{2}$, defined as the \textit{stillness score}, characterizes movement stability during food transport towards the mouth. Following identification of the upward movement sequence between food gathering and mouth insertion, the corresponding IMU signals are isolated and analyzed. Observations from our dataset, indicate that increased utensil load manifests through more controlled and deliberate movements. We capture this behavior through two key indicators; the movement's stability, quantified by linear normalization (min-max) of the mean signal variance $V_{norm}$ across all six IMU axes (experimentally bounded to $[1, 10]$), and temporal control measured by linear normalization of the movement's duration $D_{norm}$ (bounded to $[0, 35]$ samples \cite{c10}). Through experimentation of various mathematical combinations of these indicators, the formula $(1 - V_{norm}) + \log(D_{norm} + 1)$ was empirically determined to best describe the movement's steadiness, thus creating feature  $f_{2}$.

\subsubsection{Statistical Features}

Following \cite{c9}, features $f_{3}$ through $f_{6}$ are derived from sliding window analysis of the data matrix $\mathbf{M}$ (window size $2$s, step $0.1$s). For each window, we compute four intermediate features: gyroscope energy $E$ computed as the sum of squared values across the three gyroscope axes [$\mathbf{g}_x, \mathbf{g}_y, \mathbf{g}_z$], total variance $V$ across all signals from $\mathbf{M}$, y-axis gyroscope range $R$ and z-axis acceleration entropy $S$. The final features are obtained through statistical aggregation of all windows: skewness of $E$ and $V$ (features $f_{3}$ and $f_{4}$), maximum of $R$ (feature $f_{5}$), and minimum of $S$ (feature $f_{6}$), thus creating the final feature vector $\mathbf{F} = [f_{1}, f_{2}, ..., f_{6}] \in \mathbf{R}^6$.

\subsection{Bite weight estimation model}

Through systematic evaluation of traditional and deep learning approaches, a Support Vector Regression (SVR) model was selected for its optimal performance and low complexity. The SVR, implements a linear mapping $f: \mathbf{R}^6 \rightarrow \mathbf{R}$ in the original feature space, determined through optimization of the regularized risk functional with $\varepsilon$-insensitive loss. An extensive grid search across model hyperparameters, yielded optimal values of $C = 1.01$ and $\varepsilon = 0.016$. The model incorporates feature standardization (z-score), ensuring balanced contribution during training. Figure \ref{fig:algorithm_pipeline} visualizes the proposed method's pipeline.

\subsection{Comparative architectures}

For comprehensive evaluation, we implemented two comparison approaches. First, an adapted version of Mirtchouk \textit{et al.}'s \cite{c9} multimodal approach. While the original method combines IMU data from multiple body locations with audio and annotated features, our adaptation maintains core methodology while operating exclusively on dominant wrist IMU data. Feature extraction follows their sliding window technique ($5$s window duration with $0.1$s step), computing 11 statistical features (mean, covariance, and derivatives), 15 temporal shape features (coefficients of polynomial fit to acceleration components), and two frequency domain features (zero-crossing rate and its standard deviation). Subsequently, mean and standard deviation of these features are computed across the window frames within each bite, yielding their final feature vector. The random forest regression model preserves original configuration (40 trees).

For our second comparison, we implemented an early fusion architecture, inspired by \cite{c12}, to explore the potential benefits of combining IMU signals with engineered features in a trainable deep learning architecture. All architectural and training hyperparameters were determined through repeated five-fold cross-validation with five iterations to mitigate random effects. The architecture processes post-padded IMU sequences $\mathbf{M}$ of max length $t=12$s (selected based on the 95th percentile of observed bite durations) through two LSTM layers (32 with returned sequences and 16 units respectively). A fusion layer then concatenates the LSTM-processed IMU signals with our feature vector $\mathbf{F}$, forming a joint representation that passes through two dense layers (32, 8 units) with ReLu activation and an intermediate 20\% dropout, culminating in a linear output neuron. Training was performed for 300 epochs, using Adam optimizer with learning rate $10^{-2}$, batch size 64 and mean squared error as the loss function.

\section{EVALUATION}

\subsection{Dataset}

\begin{table}[b]
\centering
\caption{Collected dataset information.}
\label{tab:annotated_stats}
    \begin{tabular}{l r}
    \toprule
    \textbf{Metric} & \textbf{Value} \\
    \midrule
    Number of sessions & 10 \\
    Total Bites & 342 \\
    Mean session duration (std) & 23.42 (12.15) min \\
    Mean bite duration (std) & 6.45 (3.41) sec \\
    Bite duration range & 1.61 sec -- 27.19 sec \\
    Mean bite weight (std) & 10.89 (6.04) g \\
    Bite weight range & 0.0 g -- 34.0 g \\
    \bottomrule
    \end{tabular}
\end{table}

To evaluate the proposed approach, we collected an in-house dataset (Table \ref{tab:annotated_stats}) comprising ten participants (20\% female, age 24 $\pm$ 0.77 years) during semi-controlled eating sessions. Participants selected their meals with the constraint of using conventional utensils (fork, spook, knife) and single-plate presentation, to enable weight measurement. The dataset  encompasses 3.9 hours of recording across sessions.

Inertial signals were captured using a commercial smartwatch (Huawei Watch 2) worn on the dominant wrist, via custom Android applications, recording synchronized accelerometer and gyroscope signals. Ground truth bite weights were measured using a Bluetooth-enabled plate scale \cite{c13}, sampled at $1$ Hz. Video recordings (GoPro Hero $10$) were used to establish ground truth for the temporal segmentation of bite events. Each bite's start and end timestamps were manually annotated through frame-by-frame video analysis. Data source synchronization was achieved through three distinct hand claps detectable by the annotator in both the video and inertial signal streams. Manual video annotation yielded a total of 342 bites, with corresponding weight measurements derived from the plate scale recordings. The dataset is available upon request from the authors.

\subsection{Framework}

Evaluation follows the LOSO scheme, ensuring full separation between training and test data at the subject level. For objective performance assessment, we establish a reference point through a baseline model. This statistical baseline represents the simplest reasonable predictor - one that always predicts the mean bite weight observed in the training data. The prediction for each bite $i$ of subject $s$, is given by $\hat{y}_{s,i} = \frac{1}{N_{s}} \sum_{j \in B_{s}} y_j$, where $N_s$ is the number of bites from all participants except s, $B_s$ their set, and $y_j$ the weight of bite $j$. The metric \textit{improvement}, quantifies relative performance difference over this baseline, through mean absolute error (MAE), defined as $\left(\text{MAE}_{\text{baseline}} - \text{MAE}_{\text{model}}\right) / \text{MAE}_{\text{baseline}} \times 100$. This relative metric contextualizes model performance, beyond standalone error metrics. Finally, each model was trained based on its specific preprocessing requirements and evaluated on a common subset based on the intersection of all preprocessed subsets.

\section{RESULTS}

\begin{figure}[t]
    \centering
    \includegraphics[width=0.95\columnwidth]{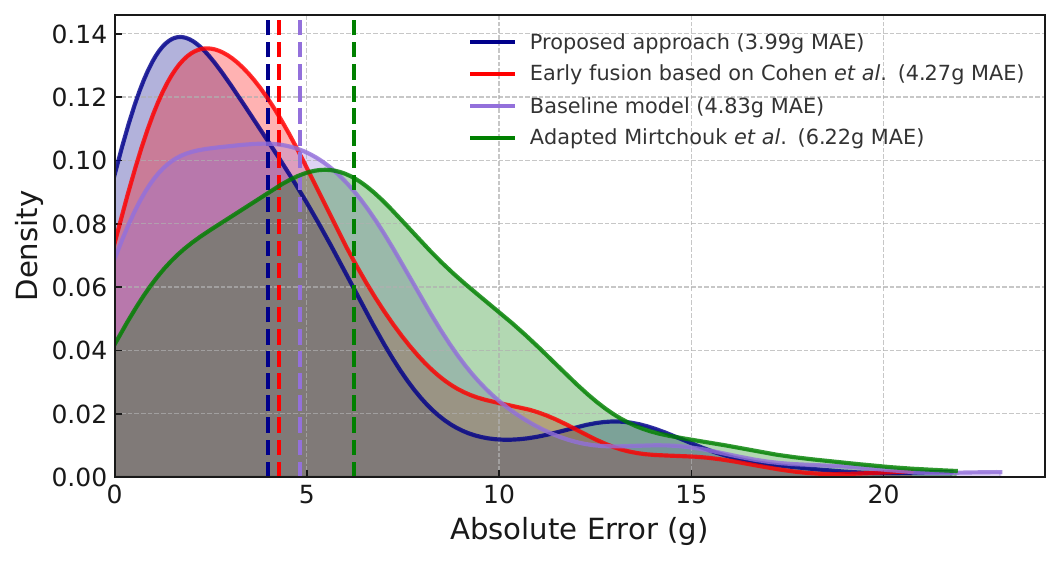}
    \caption{Figure illustrating the absolute error distribution across evaluated methods. The dotted vertical line represents the MAE.}
    \label{fig:ae_comparison}
\end{figure}

Table \ref{tab:model_performance} presents comparative performance across architectures under LOSO CV. The proposed SVR model achieves superior performance with MAE of $3.99$g per bite, a $17.41\%$ improvement over the baseline. Early fusion demonstrates moderate success with an $11.59\%$ improvement (MAE: $4.27$g). Notably, this architecture achieves the lowest MSE ($30.84$g$^2$), suggesting enhanced capability in capturing outlier movements through direct IMU processing, though its higher MAPE ($48.44\%$) indicates this additional information does not consistently improve weight estimation. The adapted Mirtchouk \textit{et al.} \cite{c9} algorithm achieves MAE of $6.22$g, underperforming the baseline ($-28.89\%$). While their original work was designed around richer multimodal input, extracting meaningful weight estimators from a single device poses distinct challenges, as reflected in these results. Figure \ref{fig:ae_comparison} shows this gap via absolute error distributions, where the proposed approach has more concentrated errors, faster decay, and a pronounced peak near zero, indicating more reliable estimations.

Subject-level analysis demonstrates the proposed approach's capabilities in adapting across different eating sessions. Individual improvements ranged from $55.42\%$ (MAE: $2.39$g) to $-33.55\%$ (MAE: $2.56$g). When analyzing total meal weight, we observe a somewhat consistent behavior, where the mean total meal difference is at $-20.04$g, with a range of $-4.8$g to $-177.9$g, in which latter case the improvement against the baseline model remained positive at $6.48\%$ denoting the model's adaptability even in cases of irregular eating patterns.

\begin{table}[b]
\centering
\caption{Comparative performance results under LOSO. N/A indicates cases where the improvement metric is inapplicable, as it measures relative performance against the baseline model.}
\label{tab:model_performance}
    \resizebox{\columnwidth}{!}{%
        \begin{tabular}{l c c c c c}
        \toprule
        \textbf{Model} & \textbf{Improvement (\%)} & \textbf{MAE (g)} & \textbf{MAPE (\%)} & \textbf{MSE (g$^{2}$)} \\
        \midrule
        Proposed approach & \textbf{17.41} & \textbf{3.99} & \textbf{40.10} & 31.45 \\
        Early fusion (based on \cite{c12})  & 11.59 & 4.27 & 48.44 & \textbf{30.84} \\
        Baseline & N/A & 4.83 & 56.66 & 39.02 \\
        Adapted Mirtchouk \textit{et al.} \cite{c9} & -28.89 & 6.22 & 74.77 & 56.02 \\
        \bottomrule
        \end{tabular}%
    }
\end{table}

\section{CONCLUSIONS}

This work presents a novel approach to estimating food intake quantity using only inertial signals from commercial smartwatches, showcasing that wrist motion encodes weight-related information. Weight estimation with a single device marks a significant step toward practical dietary monitoring. While our approach relies on manual bite annotations, its accuracy provides a strong foundation for integration with automated meal and bite detection \cite{c10, c11}. To facilitate research, our dataset is available upon request. Future work includes evaluation on larger, more diverse datasets and integration with automated detection for real-world deployment of accessible, non-invasive monitoring.

\addtolength{\textheight}{-12cm}   




\section*{ACKNOWLEDGMENT}

This work was partially supported by REBECCA (European Union’s Horizon 2020 research and innovation programme under grant agreement No 965231).


\bibliographystyle{IEEEtran} \bibliography{root}

\end{document}